\documentclass[10pt,letterpaper,twoside,british,10pt]{article}
\usepackage[T1]{fontenc}
\usepackage[latin9]{inputenc}
\usepackage{amsmath}
\usepackage{amssymb}
\usepackage{graphicx}

\makeatletter

\usepackage{spconf}
\DeclareMathOperator{\diag}{diag}
\usepackage{url}

\makeatother

\usepackage{babel}
\begin{document}
\name{Jean-Marc Valin and Iain B. Collings
\thanks{\copyright 2007 IEEE.  Personal use of this material is permitted. Permission from IEEE must be obtained for all other uses, in any current or future media, including reprinting/republishing this material for advertising or promotional purposes, creating new collective works, for resale or redistribution to servers or lists, or reuse of any copyrighted component of this work in other works.}}\address{\vspace{-0.18cm}CSIRO ICT Centre, Sydney, Australia (jmvalin@jmvalin.ca)}

\title{A New Robust Frequency Domain Echo Canceller With Closed-Loop Learning
Rate Adaptation}
\maketitle
\begin{abstract}
One of the main difficulties in echo cancellation is the fact that
the learning rate needs to vary according to conditions such as double-talk
and echo path change. Several methods have been proposed to vary the
learning. In this paper we propose a new closed-loop method where
the learning rate is proportional to a misalignment parameter, which
is in turn estimated based on a gradient adaptive approach. The method
is presented in the context of a multidelay block frequency domain
(MDF) echo canceller. We demonstrate that the proposed algorithm outperforms
current popular double-talk detection techniques by up to 6 dB.
\end{abstract}

\section{Introduction}

In any echo cancellation system, the presence of near end speech (double-talk)
tends to make the adaptive filter diverge. To counter the effect,
robust echo cancellers require adjustment of the learning rate to
account for the presence of double-talk in the signal.

Most echo cancellation algorithms attempt to explicitly detect double-talk
\cite{Benesty2000} conditions and then react by freezing the coefficients
of the adaptive filter (setting the learning rate to zero). Reliable
double-talk detection is a difficult problem and sometimes it is not
clear what should be considered as double-talk, especially in an acoustic
echo cancellation context with stationary background noise.

In previous work \cite{ValinAEC2006}, we proposed a new approach
to make echo cancellation more robust to double-talk. Instead of attempting
to explicitly detect double-talk conditions, a continuous learning
rate was used. The learning rate depends on a misalignment estimate,
which is obtained through a linear regression. While the technique
gives good results, the estimation of the misalignment remains a difficult
problem. 

In this paper, we propose a new approach where the misalignment is
estimated in closed-loop based on a gradient adaptive approach. This
closed-loop technique is applied to the block frequency domain (MDF)
adaptive filter \cite{Soo1990} and shows a significant improvement
over previous approaches.

In Section \ref{sec:Optimal-NLMS-Adaptation}, we discuss the optimal
learning rate in presence of noise. Section \ref{sec:Application-to-MDF}
introduces the multidelay frequency domain (MDF) adaptive filter.
In Section \ref{sec:Gradient-Adaptive-Learning-Rate}, we propose
a gradient adaptive technique for adjusting the learning rate of the
MDF algorithm. Experimental results and a discussion are presented
in Section \ref{sec:Results} and Section \ref{sec:Conclusion} concludes
this paper.

\section{Optimal Learning Rate In the Presence of Double-Talk}

\label{sec:Optimal-NLMS-Adaptation}

\begin{figure}
\begin{center}\includegraphics[width=0.9\columnwidth,keepaspectratio]{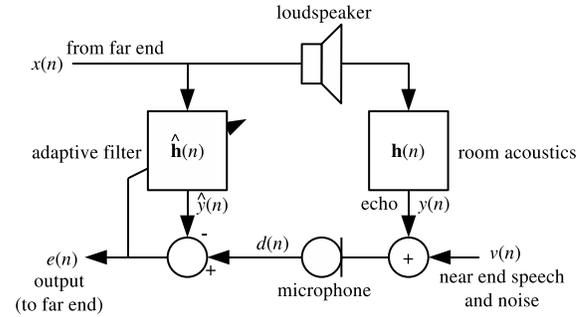}\end{center}

\caption{Block diagram of echo cancellation system.\label{cap:Block-diagram}}
\end{figure}

In the acoustic echo cancellation context illustrated in Fig. \ref{cap:Block-diagram},
the speech signal $x\left(n\right)$ received from the far end is
played into a loudspeaker. The microphone signal $d\left(n\right)$
captures both the convoluted far end speech $y\left(n\right)$ and
the near end noisy speech $v\left(n\right)$. The adaptive filter
attempts to estimate the impulse response $\mathbf{\hat{h}}\left(n\right)$
to be as close as possible to the real impulse response $\mathbf{h}\left(n\right)$.
The estimated echo $\hat{y}\left(n\right)$ is subtracted from $d\left(n\right)$,
so the output signal $e\left(n\right)$ contains both double-talk
$v\left(n\right)$ and residual echo $r\left(n\right)=y\left(n\right)-\hat{y}\left(n\right)$.

The conventional approach to double-talk robustness consists of setting
the learning rate to zero when double-talk is detected. Double-talk
detectors \cite{Benesty2000} are thus an important aspect of the
approach. Unfortunately, they are sometimes unreliable, especially
in acoustic echo cancellation context when background noise is present.
In this paper, we investigate continuous learning rates that do not
depend on a binary double-talk decision.

Whenever an adaptive filter is not perfectly adjusted, its residual
signal $r\left(n\right)$ can be used to gain better information about
the exact (time-varying) filter weights $\mathbf{h}\left(n\right)$.
However, the amount of information about $\mathbf{h}\left(n\right)$
present in $e\left(n\right)$ decreases with the amount of noise and
near end speech $v(n)$. In the case of the normalised least mean
square (NLMS) filter, it means that the stochastic gradient becomes
less reliable when the noise increases or when the filter misalignment
decreases (as the filter converges). The theoretical optimal learning
rate is approximately equal to the residual-to-error ratio \cite{ValinAEC2006}:
\begin{equation}
\mu_{opt}(n)\approx\frac{E\left\{ r^{2}\left(n\right)\right\} }{E\left\{ e^{2}\left(n\right)\right\} }\label{eq:Residual-to-output-rate}
\end{equation}
where $r\left(n\right)=y\left(n\right)-\hat{y}\left(n\right)$ is
the (unknown) residual echo and $e\left(n\right)$ is the error signal.

One possible method to vary the learning rate $\mu\left(n\right)$
would be to use the generalized normalized gradient descent (GNGD)
algorithm \cite{Mandic2004}, which includes the NLMS learning rate:
\begin{equation}
\mu\left(n\right)=\frac{\mu_{0}E\left\{ x^{2}\left(n\right)\right\} }{E\left\{ x^{2}\left(n\right)\right\} +\epsilon\left(n\right)}\label{eq:GNGD}
\end{equation}
where $\epsilon\left(n\right)$ is adapted based on the NLMS stochastic
gradient behaviour. To examine $\epsilon\left(n\right)$ more closely,
it is reasonable to surmise that (\ref{eq:GNGD}) eventually converges
to the optimal learning rate defined by (\ref{eq:Residual-to-output-rate}).
Assuming steady state behaviour ($\epsilon\left(n\right)$ is stable)
and $\mu_{0}=1$, we find that:
\begin{equation}
\frac{E\left\{ r^{2}\left(n\right)\right\} }{E\left\{ r^{2}\left(n\right)\right\} +\gamma\left(n\right)\epsilon\left(n\right)}\approx\frac{E\left\{ r^{2}\left(n\right)\right\} }{E\left\{ e^{2}\left(n\right)\right\} }
\end{equation}
where $\gamma=E\left\{ r^{2}\left(n\right)\right\} /E\left\{ x^{2}\left(n\right)\right\} $.
Knowing that we have $E\left\{ e^{2}\left(n\right)\right\} =E\left\{ r^{2}\left(n\right)\right\} +E\left\{ v^{2}\left(n\right)\right\} $,
we find the relation $\epsilon\left(n\right)\approx E\left\{ v^{2}\left(n\right)\right\} /\gamma\left(n\right)$.
In other words, the gradient-adaptive parameter $\epsilon\left(n\right)$
is approximately proportional to the variance of the near-field signal
and independent of the far-field signal. Because $\epsilon\left(n\right)$
can only be adapted slowly over time, it is clear that (\ref{eq:GNGD})
implicitly assumes that $E\left\{ v^{2}\left(n\right)\right\} $ also
varies slowly. While this is a reasonable assumption in applications
where the background noise is constant or slowly varying, the assumption
does not hold for acoustic echo cancellation, where double-talk can
start or stop at any time.

In previous work \cite{ValinAEC2006}, we proposed to use (\ref{eq:Residual-to-output-rate})
directly to adapt the learning rate. While $E\left\{ e^{2}\left(n\right)\right\} $
can easily be estimated, the estimation of the residual echo $E\left\{ r^{2}\left(n\right)\right\} $
is difficult because one does not have access to the real filter coefficients.
One reasonable assumption we make is that: 
\begin{equation}
E\left\{ r^{2}\left(n\right)\right\} =\eta\left(n\right)E\left\{ y^{2}\left(n\right)\right\} \approx\eta\left(n\right)E\left\{ \hat{y}^{2}\left(n\right)\right\} 
\end{equation}
where $\eta\left(n\right)$ is the normalised filter misalignment
(or the inverse of the echo return loss enancement) and is easier
to estimate because it is assumed to vary slowly as a function of
time. Although direct estimation of $\eta\left(n\right)$ through
linear regression can lead to good results, estimating $\eta\left(n\right)$
remains a difficult problem. In this paper we propose to apply a gradient
adaptive approach to the problem of estimating $\eta\left(n\right)$.

\section{The MDF Algorithm}

\label{sec:Application-to-MDF}

In this paper, we consider the special case of the multidelay block
frequency domain (MDF) adaptive filter \cite{Soo1990}. The MDF algorithm
in matrix form is detailed here for the sake of clarity. Let $N$
be the MDF block size, $K$ be the number of blocks and $\mathbf{F}$
denote the $2N\times2N$ Fourier transform matrix, we denote the frequency-domain
signals for frame $\ell$ as:
\begin{align}
\underline{\mathbf{e}}\left(\ell\right) & \!=\!\mathbf{F}\left[\mathbf{0}_{1\times N},e\left(\ell N\right),\ldots,e\left(\ell N+N-1\right)\right]^{T}\\
\underline{\mathbf{x}}_{k}\!\!\left(\ell\right) & \!=\!\diag\!\left\{ \!\mathbf{F}\!\left[x\!\left((\ell{-}k{-}1)N\right)\!,\!\ldots,\!x\!\left((\ell{-}k{+}1)N{-}1\right)\right]^{T}\!\right\} \\
\underline{\mathbf{X}}\left(\ell\right) & \!=\!\left[\underline{\mathbf{x}}_{0},\underline{\mathbf{x}}_{1},\ldots,\underline{\mathbf{x}}_{K-1}\right]\\
\underline{\mathbf{d}}\left(\ell\right) & \!=\!\mathbf{F}\left[\mathbf{0}_{1\times N},d\left(\ell N\right),\ldots,d\left(\ell N+N-1\right)\right]^{T}
\end{align}
The MDF algorithm is then expressed in matrix form as:
\begin{align}
\underline{\mathbf{e}}\left(\ell\right) & =\underline{\mathbf{d}}\left(\ell\right)-\underline{\mathbf{y}}\left(\ell\right)\\
\underline{\hat{\mathbf{y}}}\left(\ell\right) & =\mathbf{G}_{1}\underline{\mathbf{X}}\left(\ell\right)\mathbf{\underline{\hat{h}}}\left(\ell\right)\\
\mathbf{\underline{\hat{h}}}\left(\ell+1\right) & =\mathbf{\underline{\hat{h}}}\left(\ell\right)+\mathbf{G}_{2}\underline{\boldsymbol{\mu}}\left(\ell\right)\nabla\mathbf{\hat{\underline{h}}}\left(\ell\right)\\
\nabla\mathbf{\underline{\hat{h}}}\left(\ell\right) & =\boldsymbol{\Phi}_{\mathbf{xx}}^{-1}\left(\ell\right)\underline{\mathbf{X}}^{H}\left(\ell\right)\mathbf{\underline{e}}\left(\ell\right)
\end{align}
where $\boldsymbol{\Phi}_{\mathbf{xx}}\left(\ell\right)$ is the diagonal
normalisation matrix as computed in \cite{Soo1990}, $\mathbf{G}_{1}$
and $\mathbf{G}_{2}$ are the constraint matrices:
\begin{align}
\mathbf{G}_{1} & =\mathbf{F}\left[\begin{array}{cc}
\mathbf{0}_{N\times N} & \mathbf{0}_{N\times N}\\
\mathbf{0}_{N\times N} & \mathbf{I}_{N\times N}
\end{array}\right]\mathbf{F}^{-1}\\
\tilde{\mathbf{G}}_{2} & =\mathbf{F}\left[\begin{array}{cc}
\mathbf{I}_{N\times N} & \mathbf{0}_{N\times N}\\
\mathbf{0}_{N\times N} & \mathbf{0}_{N\times N}
\end{array}\right]\mathbf{F}^{-1}\\
\mathbf{G}_{2} & =\textrm{diag}\left\{ \tilde{\mathbf{G}}_{2},\tilde{\mathbf{G}}_{2},\ldots,\tilde{\mathbf{G}}_{2}\right\} 
\end{align}
 and $\underline{\boldsymbol{\mu}}\left(\ell\right)$ is the $2KN\times2KN$
diagonal learning rate matrix:
\begin{align}
\underline{\boldsymbol{\mu}}\left(\ell\right) & =\diag\left\{ \left[\boldsymbol{\mu}^{T}\left(\ell\right),\boldsymbol{\mu}^{T}\left(\ell\right),\ldots,\boldsymbol{\mu}^{T}\left(\ell\right)\right]^{T}\right\} \\
\boldsymbol{\mu}\left(\ell\right) & =\left[\mu_{0}\left(\ell\right),\mu_{1}\left(\ell\right),\ldots,\mu_{2N-1}\left(\ell\right)\right]^{T}
\end{align}
If $\mu_{k}\left(\ell\right)=\mu_{o}$, we have the standard MDF algorithm.

\section{Gradient-Adaptive Learning Rate}

\label{sec:Gradient-Adaptive-Learning-Rate}

In \cite{ValinAEC2006}, we proposed to use the frequency-dependent
learning rate:
\begin{equation}
\mu_{k}\left(\ell\right)=\min\left(\eta\left(\ell\right)\frac{\left|\underline{\mathbf{\hat{y}}}_{k}\left(\ell\right)\right|^{2}}{\left|\underline{\mathbf{e}}_{k}\left(\ell\right)\right|^{2}},\mu_{0}\right)\label{eq:freq-learning-rate}
\end{equation}
where $\eta(\ell)$ is a direct estimation of the normalised misalignment
exploiting the non-stationarity of the signals and using linear regression
between the power spectra of the estimated echo and the output signal.
The motivation behind this formulation is that it factors the residual
echo estimation into a slowly-evolving (but unfortunately difficult
to estimate) normalised misalignment $\eta(\ell)$ and a rapidly-evolving
(but easy to estimate) term far-end term $\left|\underline{\mathbf{y}}_{k}\left(\ell\right)\right|^{2}$.
The learning rate can thus react quickly to double-talk even if the
estimation of the residual echo (through the misalignment estimate)
requires a longer time period. A remaining problem with that approach
is that $\eta(\ell)$ is difficult to estimate and the algorithm does
not know whether its estimate of $\eta(\ell)$ is too low or too high.
In this sense, the $\eta(\ell)$ update in \cite{ValinAEC2006} is
an open-loop estimate.

\subsection{Adaptation algorithm}

In this paper we bypass the difficulty of estimating $\eta(\ell)$
directly and instead propose a closed-loop gradient adaptive estimation
of $\eta(\ell)$. The parameter $\eta(\ell)$ is no longer an estimate
of the normalised misalignment, but is instead adapted in closed-loop
in such a way as to maximise convergence of the adaptive filter. As
with other gradient-adaptive methods \cite{Mathews1993,Mandic2004}
we compute the derivative of the mean square error $E\left(\ell\right)=\underline{\mathbf{e}}^{H}\left(\ell\right)\underline{\mathbf{e}}\left(\ell\right)$,
this time with respect to $\eta\left(\ell-1\right)$, using the chain
derivation rule:
\begin{align}
\frac{\partial E\left(\ell\right)}{\partial\eta\!\left(\ell{-}1\right)} & \!=\frac{\partial E\left(\ell\right)}{\partial\underline{\mathbf{e}}\left(\ell\right)}\frac{\partial\underline{\mathbf{e}}\left(\ell\right)}{\partial\underline{\mathbf{\hat{y}}}\left(\ell\right)}\frac{\partial\underline{\mathbf{\hat{y}}}\left(\ell\right)}{\partial\underline{\hat{\mathbf{h}}}\left(\ell\right)}\frac{\partial\underline{\hat{\mathbf{h}}}\left(\ell\right)}{\partial\mu\left(\ell-1\right)}\frac{\partial\mu\left(\ell-1\right)}{\partial\eta\left(\ell-1\right)}\\
 & \!={-}\underline{\mathbf{e}}^{H}\!\left(\ell\right)\!\mathbf{G}_{1}\underline{\mathbf{X}}\!\left(\ell\right)\!\mathbf{G}_{2}\boldsymbol{\Phi}_{\mathbf{xx}}^{-1}\!\left(\ell\right)\!\underline{\mathbf{X}}^{H}\!\left(\ell\right)\mathbf{\underline{e}}\!\left(\ell\right)\!\frac{\left|\underline{\mathbf{\hat{y}}}_{k}\!\left(\ell\right)\right|^{2}}{\left|\underline{\mathbf{e}}_{k}\!\left(\ell\right)\right|^{2}}\\
 & \!\approx{-}\frac{\left|\underline{\mathbf{\hat{y}}}_{k}\left(\ell\right)\right|^{2}}{\left|\underline{\mathbf{e}}_{k}\left(\ell\right)\right|^{2}}\nabla\mathbf{\underline{\hat{h}}}^{H}\left(\ell\right)\boldsymbol{\Phi}_{\mathbf{xx}}\nabla\mathbf{\underline{\hat{h}}}\left(\ell\right)
\end{align}
We propose to use a filtered version of the gradient with a multiplicative
update, similar to the general approch in \cite{Ang2001}, and to
drop the normalisation term $\boldsymbol{\Phi}_{\mathbf{xx}}$. This
results in the new gradient-based learning rate adaptation rule: 
\begin{align}
\eta\left(\ell+1\right) & =\eta\left(\ell\right)\exp\!\left[\rho\frac{\left|\underline{\mathbf{\hat{y}}}_{k}\left(\ell\right)\right|^{2}\Re\left\{ \underline{\boldsymbol{\psi}}^{H}\left(\ell\right)\nabla\mathbf{\hat{\underline{h}}}\left(\ell\right)\right\} }{\left|\underline{\mathbf{e}}_{k}\left(\ell\right)\right|^{2}\sum_{k}\!\left|\Re\!\left\{ \psi_{k}^{*}\left(\ell\right)\nabla\hat{h}_{k}\left(\ell\right)\right\} \right|}\right]\label{eq:gradient-adaptation}\\
\underline{\boldsymbol{\psi}}\left(\ell+1\right) & =\alpha\underline{\boldsymbol{\psi}}\left(\ell\right)+\nabla\mathbf{\underline{\hat{h}}}\left(\ell\right)\label{eq:gradient-smoothing}
\end{align}
where $\rho$ is the step size and $\alpha$ controls the gradient
smoothing (typically $\rho=1$, $\alpha=0.9$). 

An intuitive interpretation for the update equation in (\ref{eq:gradient-adaptation})
is that when the learning rate is too high (because $\eta\left(\ell\right)$is
too high), the gradient oscillates, causing $\eta\left(\ell\right)$
to decrease. On the other hand, when the learning rate is too low
(perhaps because the echo path has changed) the gradient keeps pointing
in the same direction and $\eta\left(\ell\right)$ increases. Also,
because of the $\frac{\left|\underline{\mathbf{\hat{y}}}_{k}\left(\ell\right)\right|^{2}}{\left|\underline{\mathbf{e}}_{k}\left(\ell\right)\right|^{2}}$
factor, we can apply (\ref{eq:gradient-adaptation}) at each frame
regardless of double-talk or even when no far-end speech is present
(in which case $\left|\underline{\mathbf{\hat{y}}}_{k}\left(\ell\right)\right|^{2}=0$
anyway). 

The last aspect that needs to be addressed is the initial condition.
When the filter is initialised, all the weights are set to zero, which
means that $\underline{\mathbf{\hat{y}}}\left(\ell\right)=\mathbf{0}$
and no adaptation can take place in (\ref{eq:freq-learning-rate})
and (\ref{eq:gradient-adaptation}). In order to bootstrap the adaptation
process, the learning rate $\mu_{k}\left(\ell\right)$ is set to a
fixed constant (we use $\mu_{k}\left(\ell\right)=0.25$) for a short
time equal to twice the filter length (only non-zero portions of signal
$x(n)$ are taken into account). This procedure is only necessary
when the filter is initialised and is not required in case of echo
path change.

\subsection{Algorithm behaviour}

Here we show that the adaptive learning rate described above is able
to deal with both double-talk and echo path change without explicit
modelling. From (\ref{eq:freq-learning-rate}) we can see that when
double-talk occurs, the denominator $\left|\underline{\mathbf{e}}_{k}\left(\ell\right)\right|^{2}$
rapidly increases, causing an instantaneous decrease in the learning
rate that lasts only as long as the double-talk period lasts. In the
case of background noise, the learning rate depends on both the presence
of an echo signal as well as the misalignment estimate. As the filter
misalignment becomes smaller, the learning rate also becomes smaller. 

One major difficulty involved in double-talk detection is the need
to distinguish between double-talk and echo path change since both
cause a sudden increase in the filter error signal. This distinction
is made implicitly in the gradient-based adaptation of $\eta(\ell)$.
When the echo path changes, $\nabla\mathbf{\underline{\hat{h}}}\left(\ell\right)$
starts pointing steadily in the same direction, thus significantly
increasing $\eta(\ell)$, which is a clear sign that the filter is
no longer properly adapted.

In gradient adaptive methods \cite{Mathews1993,Ang2001}, the implicit
assumption is that both the near-end and the far-end signals are nearly
stationary. We have shown that the GNGD algorithm \cite{Mandic2004}
only requires the near-end signal to be nearly stationary. In our
new proposed method, both signals can be non-stationary, which is
a requirement for double-talk robustness.

\section{Results And Discussion}

\label{sec:Results}

The proposed system is evaluated in an acoustic echo cancellation
context with background noise, double-talk and a change in the echo
path (16 seconds into a 32-second recording). The two impulse responses
used are 1024-sample long and measured from real recordings in a small
office room with both the microphone and the loudspeaker resting on
a desk. 

The proposed algorithm is compared to our previous work \cite{ValinAEC2006},
to the normalised cross-correlation (NCC) method \cite{Benesty2000}
and to a baseline with no double-talk detection (no DTD). The optimal
threshold found for the NCC algorithm was 0.35 with a learning rate
$\mu=0.25$. It was found that choosing $\mu_{max}=0.75$ as the upper
bound on the learning rate gave good results for the proposed algorithm.
In practise, $\mu_{max}$ has little impact on the algorithm because
the gradient-based adaptation would compensate for a higher value
of $\mu_{max}$ by reducing $\eta\left(\ell\right)$. The GNGD algorithm
\cite{Mandic2004} is not included in the evaluation because it is
not intended for applications where the near-field signal is highly
non-stationary and it was not possible to obtain better results than
the baseline. 

\begin{figure}[t]
\begin{center}\includegraphics[width=1\columnwidth,keepaspectratio]{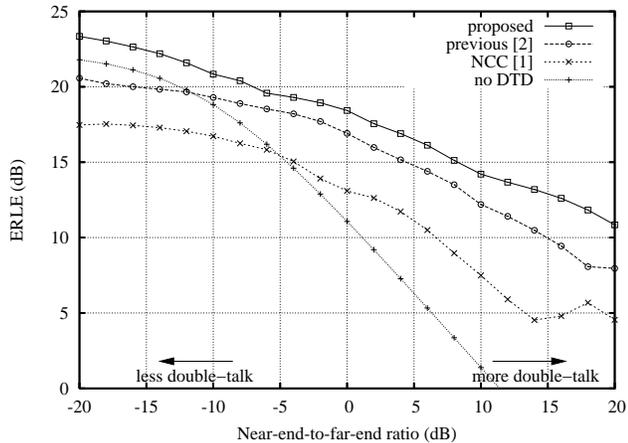}\end{center}

\caption{Steady-state ERLE (first two seconds of adaptation not considered)
as a function of the near-end to far-end ratio. \label{cap:ERLE-NFR}}
\end{figure}

\begin{figure}
\begin{center}\includegraphics[width=1\columnwidth,keepaspectratio]{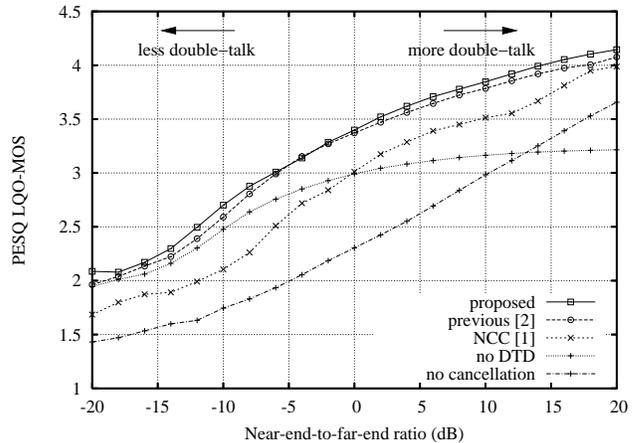}\end{center}

\caption{PESQ objective listening quality measure (LQO-MOS) as a function of
the near-end to far-end ratio.\label{cap:PESQ-NFR}}
\end{figure}

Fig. \ref{cap:ERLE-NFR} shows the average steady-state (the first
2 seconds of adaptation are not considered) ERLE for the test data
with different ratios of near-end signal and echo. Clearly, the proposed
algorithm performs better than both our previous work (2 dB average
improvement) and the NCC algorithm (6 dB average improvement). The
perceptual quality of the output speech signal is also evaluated by
comparing it to the near field signal $v(n)$ using the Perceptual
Evaluation of Speech Quality (PESQ) ITU-T recommendations P.862 \cite{P.862}.
The perceptual quality of the speech shown in Fig. \ref{cap:PESQ-NFR}
is evaluated based on the entire file, including the adaptation time.
It is again clear that the proposed algorithm performs better than
all other algorithms. It is worth noting that the reason why the results
in Fig. \ref{cap:ERLE-NFR} improve with double-talk (unlike in Fig.
\ref{cap:ERLE-NFR}) is that the signal of interest is the double-talk
$v(n)$, so the higher the double-talk the less (relative) echo in
the input signal.

\section{Conclusion}

\label{sec:Conclusion}

We have demonstrated a novel method for adjusting the learning rate
of frequency-domain adaptive filters based on a gradient adaptive
estimation of the current misalignment. The proposed method performs
better than a double-talk detector and previous work using direct
estimation of the misalignment. In addition, the closed-loop gradient
adaptive estimation of $\eta\left(\ell\right)$ makes the algorithm
conceptually simple and means that there are very few important parameters
to be tuned. Although the proposed algorithm is presented in the context
of the MDF algorithm, we believe future work could apply it to other
adaptive filtering algorithms, including the NLMS algorithm.

\bibliographystyle{IEEEbib}
\bibliography{echo}

\end{document}